\pdfoutput=1
\documentclass[sigconf]{acmart}
\usepackage{graphicx}
\usepackage{amsmath}
\usepackage{epsfig}
\usepackage{booktabs}
\usepackage{subcaption}
\usepackage{multirow}
\usepackage{multicol}
\usepackage{cleveref}
\usepackage{siunitx}
\usepackage[show]{chato-notes}

\newcommand{\spara}[1]{\smallskip\noindent{\bf #1}}

\newenvironment{squishlist}
{\begin{list}{$\bullet$}
  {\setlength{\itemsep}{0pt}
   \setlength{\parsep}{3pt}
   \setlength{\topsep}{3pt}
   \setlength{\partopsep}{0pt}
   \setlength{\leftmargin}{2em}
   \setlength{\labelwidth}{1.5em}
   \setlength{\labelsep}{0.5em} } }
{\end{list}}

\AtBeginDocument{%
  \providecommand\BibTeX{{%
    \normalfont B\kern-0.5em{\scshape i\kern-0.25em b}\kern-0.8em\TeX}}}

\copyrightyear{2021}
\acmYear{2021}
\setcopyright{acmlicensed}\acmConference[ICAIF'21]{2nd ACM International Conference on AI in Finance}{November 3--5, 2021}{Virtual Event, USA}
\acmBooktitle{2nd ACM International Conference on AI in Finance (ICAIF'21), November 3--5, 2021, Virtual Event, USA}
\acmPrice{15.00}
\acmDOI{10.1145/3490354.3494370}
\acmISBN{978-1-4503-9148-1/21/11}

\begin{document}

\title{The Evolving Causal Structure of Equity Risk Factors}

\author{Gabriele D'Acunto}
\email{gabriele.dacunto@isi.it}
\orcid{na}
\affiliation{%
  \institution{Sapienza University of Rome, Italy}
}
\affiliation{%
  \institution{ISI Foundation, Turin, Italy}
}

\author{Paolo Bajardi}
\email{paolo.bajardi@isi.it}
\orcid{na}
\affiliation{%
 \institution{ISI Foundation}
 \streetaddress{Via Chisola, 5}
 \city{Turin}
 \country{Italy}}

\author{Francesco Bonchi}
\email{francesco.bonchi@isi.it}
\orcid{na}
\affiliation{%
  \institution{ISI Foundation, Turin, Italy}
}
\affiliation{%
  \institution{Eurecat, Barcelona, Spain}
}

\author{Gianmarco De Francisci Morales}
\email{gdfm@isi.it}
\orcid{na}
\affiliation{%
 \institution{ISI Foundation}
 \streetaddress{Via Chisola, 5}
 \city{Turin}
 \country{Italy}}

\renewcommand{\shortauthors}{D'Acunto, et al.}

\begin{abstract}

In recent years, \emph{multi-factor strategies} have gained increasing popularity in the financial industry, as they allow investors to have a better understanding of the risk drivers underlying their portfolios.
Moreover, such strategies promise to promote diversification and thus limit losses in times of financial turmoil.
However, recent studies have reported a significant level of redundancy between these factors, which might enhance risk contagion among multi-factor portfolios during financial crises.
Therefore, it is of fundamental importance to better understand the relationships among factors.

Empowered by recent advances in \emph{causal structure learning} methods, this paper presents a study of the causal structure of financial risk factors and its evolution over time. In particular, the data we analyze covers \num{11} risk factors concerning the US equity market, spanning a period of 29 years at daily frequency.

Our results show a statistically significant sparsifying trend of the underlying causal structure.
However, this trend breaks down during periods of financial stress, in which we can observe a densification of the causal network driven by a growth of the out-degree of the market factor node.
Finally, we present a comparison with the analysis of factors cross-correlations, which further confirms the importance of causal analysis for gaining deeper insights in the dynamics of the factor system, particularly during economic downturns.

Our findings are especially significant from a risk-management perspective.
They link the evolution of the causal structure of equity risk factors with market volatility and a worsening macroeconomic environment, and show that, in times of financial crisis, exposure to different factors boils down to exposure to the market risk factor.

\end{abstract}

\ccsdesc[300]{Mathematics of computing~Causal networks}
\ccsdesc[300]{Applied computing~Economics}
\ccsdesc[100]{Mathematics of computing~Time series analysis}

\keywords{Causal Discovery, Structure Learning, Networks Dynamics, Risk Premia}

\maketitle \sloppy

\section{Introduction}
\label{sec:intro}

Multi-factor investing strategies have gained wide adoption during the last decade, as they allow investors to have a better understanding of the risk drivers underlying a portfolio.
Such strategies promise to promote diversification and thus limit drawdown during financial turmoils~\cite{ilmanen2012death,kremer2018risk}.
However, out of hundreds existing factors, only a small number  is truly significant in explaining the cross-section of stock returns~\cite{feng2020taming,HARVEY2021}.
Therefore, an important open question across both financial research and industry concerns factors redundancy.
In particular, the adoption by researchers of different testing approaches (e.g., panel vs. cross-sectional regression), and the presence of model selection biases as those entailed by omitted variables, may contribute to the publication of new redundant factors.
\begin{figure}[t]
    \centering
    \includegraphics[width=.45\textwidth]{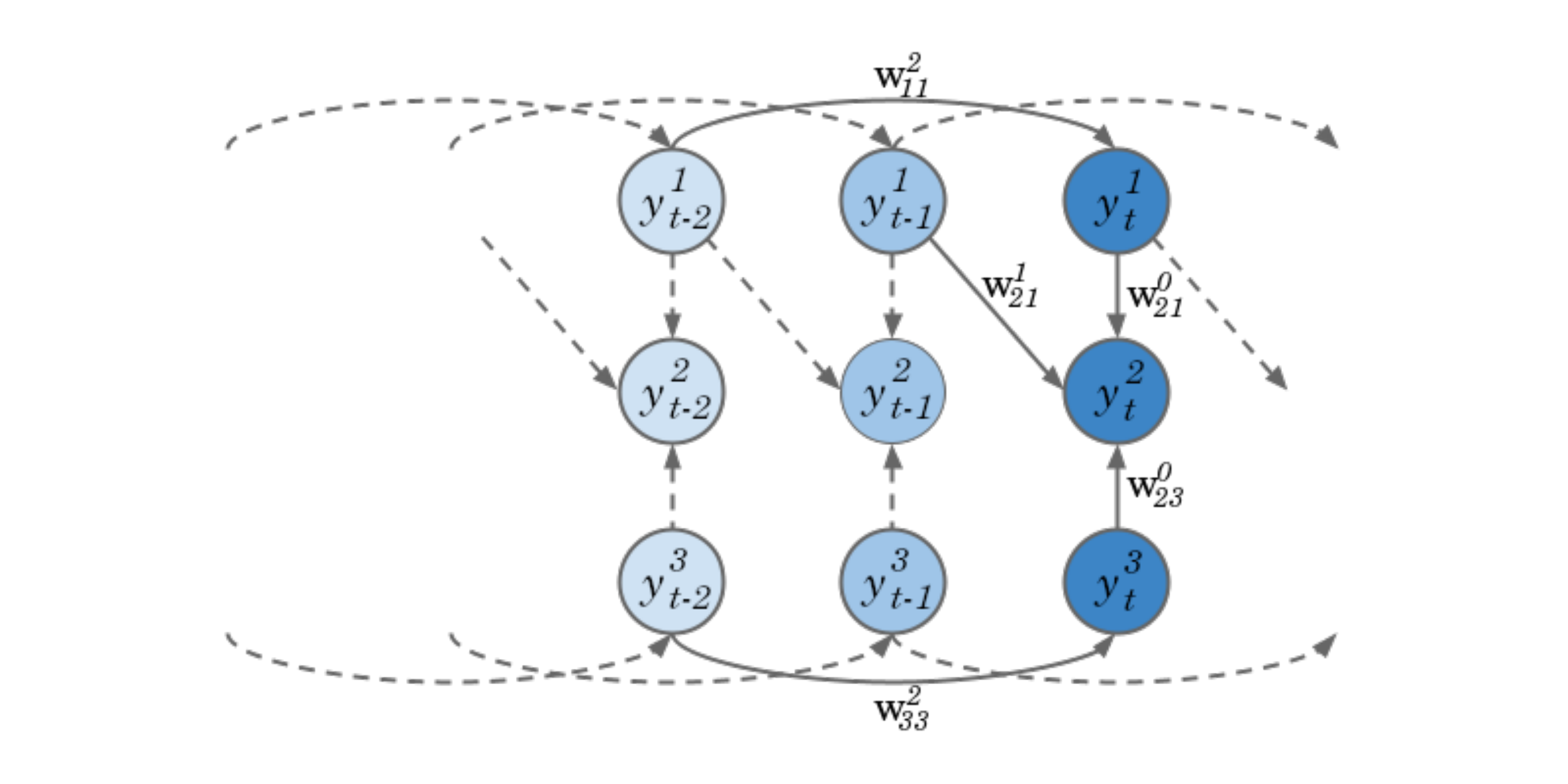}
    \vspace{-3mm}
    \caption{An example of the causal graph among equity risk factors along time with maximum lag $L=2$.}
    \label{fig:tschain}
\end{figure}
Motivated by these observations, we aim at gaining insights into the underlying dynamics of risk factor interactions by leveraging recent advances in \emph{causal structure learning}~\cite{vowels2021d}.
More precisely, starting from the known results about correlations among factors~\cite{feng2020taming,HARVEY2021}, in the spirit of Reichenbach's \emph{principle of common cause}~\cite{reich}, we investigate whether there is an underlying causal structure within the universe of considered financial factors, and how this structure evolves over time.

Specifically, our computational task is as follows:
we are given as input a dataset $\mathcal{Y} \in \mathbb{R}^{N \times T}$ composed by $N$ time series of length $T$ of factors returns, and we want to learn a graph (as the one in \Cref{fig:tschain}) representing the causal relations among the factors values along time.
More in details we are interested in understanding whether  $\mathcal{Y}$ admits a \emph{functional representation} in which each $y_t^i$ (i.e., the value of the factor $y^i$ at timestamp $t$)
is determined by at most $L$ past values of a set of other factors named \emph{parents}:
\begin{equation}\label{eq:funcrepr}
y_t^i=f^i\left((\mathcal{P}_L^i)_{t-L}, \ldots, (\mathcal{P}_0^i)_{t}, \epsilon_t^i\right), \quad i \in \{1, \ldots, N \}.
\end{equation}
Here $\epsilon_t^i$ represents the noise term, while $(\mathcal{P}_l^i)_{t-l}$ indicates the set of risk factors that cause $y_t^i$ with lag $l \in \mathbb{N}_0$.
When $l>0$, $(\mathcal{P}_l^i)_{t-l}$ may contain $y^i_{t-l}$ as well.
Conversely, as we cannot observe causal effects from present to past, $(\mathcal{P}_0^i)_{t}$ can neither contain $y_t^i$ nor $y_t^j$ if $y_t^i \in (\mathcal{P}_0^j)_{t}$, otherwise it would be impossible to define the direction of the causal relation.
More in general, we require the set of Equations~\eqref{eq:funcrepr} to be \emph{acyclic}, which means that feedback loops among variables are forbidden.
This assumption plays a key role in causal inference since it allows to set causes apart from effects.

In case the aforementioned functional forms are assumed to be linear, by using matrix notation, Equations~\eqref{eq:funcrepr} can be compactly written as a \emph{structural vector autoregressive model} (SVAR)
\begin{equation}\label{eq:SVAR}
    \mathbf{y}_{t} = \sum_{l=0}^L \mathbf{W}^{l} \mathbf{y}_{t-l} + \boldsymbol{\epsilon}_{t},
\end{equation}
where $\mathbf{y}_{t} \in \mathbb{R}^{N}$ is a column vector constituted by the observations at time $t$,
$\mathbf{W}^{l} \in \mathbb{R}^{N \times N}$ are the matrices of lagged causal effects with lag $l$ up to maximum lag $L$, such that $w^{l}_{ij} \neq 0$ iff $y_{t-l}^j \in (\mathcal{P}_{l}^i)_{t-l}$.
In addition, the matrix of instantaneous causal effects $\mathbf{W}^{0}$ respects the acyclicity requirement mentioned above.
Finally, $\boldsymbol{\epsilon}_{t} \in \mathbb{R}^{N}$ is the column vector of random disturbances at time $t$.

We remark that \Cref{eq:SVAR} should not be read as a usual system of equations, but rather as a set of functions describing how certain factors determine others.
Indeed, the model is said to be \emph{structural} since it allows to compute variables (effects) by means of linear functions of other endogenous variables (causes), by taking into account both instantaneous and lagged relations (also referred to as inter-layer connections).
In particular, the previous model can be thought of as a combination of a \emph{structural equation model} (SEM~\cite{elcainf1}) and a \emph{vector autoregressive model} (VAR~\cite{sims1980macroeconomics}).

Finally, note that \Cref{eq:SVAR} entails a \emph{directed acyclic graph} (DAG) in which there is a weighted edge from $y_{t-l}^j \in (\mathcal{P}_{l}^i)_{t-l}$ (with $l \geq 0$) to $y_{t}^i$: this is the causal graph of factors along time as depicted in \Cref{fig:tschain}.

In order to study the evolution of such non-stationary system, we estimate \Cref{eq:SVAR} by adopting a sliding window approach and performing a regression analysis on the inferred causal networks.
Our data covers \num{11} risk factors concerning the US equity market, over a period of 29 years at daily frequency. Our main results can be summarized as follows:

\begin{squishlist}

\item Causal interactions between factors exhibit a statistically significant sparsification trend along time, with anomalies in periods of financial turmoil.

\item We expose a relationship between the density of the causal networks and investor sentiment.
In particular, we show that periods of worsening sentiment and business cycle phase (see Section \ref{sec:indicators}) are associated with a densification of the causal network.
This phenomenon is driven by a growth in the out-degree of the market risk factor.

\item Finally, we conduct a comparative study between causation and correlation among factors.
Our findings highlight how causal analysis better describes the importance of the market factor among the considered risk factors.
Besides, while according to correlation analysis, the business cycle indicator is not related to the evolution of the factorial system, the study of causal structures reveals a statistically significant relationship with a $95\%$ confidence level.
\end{squishlist}

The rest of the paper is organized as follows.
\Cref{sec:relwork} provides the background about factor investing and causal discovery.
\Cref{sec:data} describes the data used to conduct the analysis and presents our methodology.
\Cref{sec:results} reports the results of our analysis.
Finally, \Cref{sec:discussion}, provides a discussion on our results together with some direction for future investigation. 
\section{Related Work}
\label{sec:relwork}
An equity risk factor is a variable able to explain the cross-section of expected stock returns, i.e., how the expected return varies among stocks.
Its significance is usually assessed via the usage of linear regression models~\cite{cochrane2009asset}.
Factor investing is well rooted in finance, in particular it dates back to the first asset pricing model, the \emph{Capital Asset Pricing Model} (CAPM), introduced by \citet{sharpe1964capital}.
CAPM looks at stock returns through the exposure to one factor, the \emph{market beta}, and introduces a precise definition of risk and how it drives expected stock returns.
Successively, \citet{fama1993common} found out that the exposure to small cap stocks (SMB) and cheaper, under-priced stocks (HML) provide two additional sources of risk not captured by CAPM.
Based on these evidences, an enormous amount of research has been produced on factor investing, and hundreds of potential factors have been proposed~\cite{cochrane2011presidential,harvey2015and,mclean2016does,hou2017replicating}.
This abundance of proposed factors has opened an important question across both financial research and industry concerning their redundancy.
As already mentioned, \citet{feng2020taming} and \citet{HARVEY2021} recently pointed out that only a small number of the existing factors is truly significant in explaining stock returns cross-section.
Our study of risk factor interactions fits into this stream of research, and in particular tackles the problem by exploiting recent advances in causal learning~\cite{pearl2009causality}.

The possibility of examining a system under interventions (i.e., counterfactual analysis) makes causal analysis a powerful tool for studying complex systems~\cite{elcainf}.
However, in many cases, the underlying causal structure is unknown, and it is not possible to carry out randomized experiments in order to study the system at hand under distribution changes.
Therefore, the interest in inferring causal structures from observational data, also known as \emph{causal structure learning}, has been significantly growing during recent years.

Causal structure learning algorithms can be classified into three main families: (i) \emph{constraint-based approaches}, which make use of conditional independence tests to establish the presence of an edge between two nodes~\cite{spirtes2000,huang2020};
(ii) \emph{score-based methods}, which use several search procedures in order to optimize a certain score function~\cite{heckerman1995,chickering2002,huang2018generalized};
(iii) \emph{structural causal models}, which express a variable at a certain node as a function of its parents~\cite{lingam,hoyer,shimizu2011directlingam,peters2014,buhlmann2014cam}.
Additionally, as highlighted in \Cref{sec:intro}, whenever we are dealing with a process that evolves over time, temporal ordering drives the causal inference procedure.
Nevertheless, the main issue concerns the estimation of instantaneous effects, which must satisfy the acyclicity requirement.
With regards to Equation (\ref{eq:SVAR}), this means that $\mathbf{W}^0$ must entail the structure of a DAG.

However, it is common to deal with non-Gaussian data.
For instance, this is the case of equity time series which show \emph{heteroscedasticity} (i.e., the variance of the stock returns varies over time) and \emph{volatility clustering} (i.e., large (small) swings in stock prices tend to group together)~\cite{bollerslev1986generalized}.

This observation implies that there is additional information, not described by the covariance matrix, that can be exploited to retrieve $\mathbf{W}^0$.
Consequently, by leveraging a non-gaussianity assumption of $\boldsymbol{\epsilon}_t$, a series of linear non-Gaussian methods to estimate the model in \Cref{eq:SVAR} have been proposed in the past years~\cite{hyvarinen,moneta2013causal}.
In particular, in this study we employ VAR-LiNGAM~\cite{hyvarinen}, as provided by the python package \texttt{lingam}\footnote{\url{https://github.com/cdt15/lingam}} made available by authors.

\emph{Econophysics}, is an interdisciplinary effort devoted at analyzing the risk contagion among financial institutions by representing the financial system as a network~\cite{bardoscia2021physics}, 
Our analysis differs from  Econophysics as we do not focus on financial institutions, and instead aim to understand the dynamics of the factorial network, which captures sources of risk largely accepted and widely used by investors.
In addition, we look for causal relationships between the observed variables by means of a machine learning causal model that, differently from existing work~\cite{billio2012econometric}, allows us to evaluate the presence of instantaneous causal effects as well. 
\section{Data and Methodology}
\label{sec:data}

In this section, we first introduce the financial risk factors and provide some information about the factor dataset $\mathcal{Y}$.
Then, we focus on the additional variables selected as indicators of changes in both investor expectation and business cycle evolution.
Finally, we present the adopted causal inference approach and introduce the \emph{generalized linear model} (GLM) regression we employ to analyze the model results.

\subsection{Financial factors}
\label{sec:factdata}
We consider \num{11} risk factors at daily frequency concerning the US equity market.
The choice to use equity factors from the latter market is driven by the greater availability of data and the higher presence of results in the financial literature that can be compared to those we obtain.
More precisely, we deal with \num{7306} daily observations spanning 29 years, from 2 January 1991 to 31 December 2019.
We include in our analysis the following published risk factors, gathered directly from the websites of authors: \emph{Excess Market Return} (Mkt-RF)~\cite{jensen1972capital}, \emph{Small Minus Big} (SMB) and \emph{High Minus Low} (HML)~\cite{fama1993common}, \emph{Momentum} (UMD)~\cite{carhart1997persistence}, \emph{HML Devil} (HML Dev)~\cite{asness2013devil}, \emph{Robust Minus Weak} (RMW) and \emph{Conservative Minus Aggressive} (CMA)~\cite{fama2015five}, \emph{HXZ Investment} (R-IA) and \emph{HXZ Profitability} (R-ROE)~\cite{hou2015digesting}, \emph{Betting Against Beta} (BAB) and \emph{Quality Minus Junk} (QMJ)~\cite{asness2019quality}.

\begin{table*}[ht]
    \centering
    \small
    \caption{Summary statistics for the analyzed factors at daily frequency. Average compounded return, volatility, risk adjusted return, and Sortino Ratio are annualised.}
    \begin{tabular}{lrrrrrrrrrrr}
    \toprule
    {} &  Mkt-RF &   SMB &   HML &   RMW &   CMA &  R-IA &  R-ROE &   BAB &  HML-dev &   UMD &   QMJ \\
    \midrule
    Avg Comp. Ret. (\%) &    7.68 &  0.67 &  2.10 &  3.89 &  2.14 &  2.23 &   5.35 &  9.43 &     0.65 &  5.17 &  4.46 \\
    Volatility (\%)     &   17.46 &  9.17 &  9.61 &  7.29 &  6.53 &  6.60 &   7.43 & 11.01 &    10.46 & 13.39 &  7.97 \\
    Risk Adj. Ret. (\%) &    0.44 &  0.07 &  0.22 &  0.53 &  0.33 &  0.34 &   0.72 &  0.86 &     0.06 &  0.39 &  0.56 \\
    Sortino Ratio (\%)  &    0.62 &  0.10 &  0.32 &  0.79 &  0.47 &  0.48 &   1.04 &  1.21 &     0.09 &  0.53 &  0.83 \\
    Skew               &   -0.15 & -0.22 &  0.43 &  0.26 & -0.44 & -0.72 &  -0.11 & -0.34 &     0.52 & -0.23 &  0.20 \\
    Kurtosis           &    8.34 &  3.91 &  9.05 &  7.71 & 11.44 & 15.92 &   5.84 & 11.72 &    11.47 & 24.57 &  8.70 \\
    1st \%-ile (\%)      &   -2.98 & -1.46 & -1.67 & -1.28 & -1.10 & -1.08 &  -1.41 & -2.20 &    -1.73 & -2.52 & -1.30 \\
    5th \%-ile (\%)      &   -1.72 & -0.91 & -0.83 & -0.66 & -0.57 & -0.57 &  -0.68 & -0.97 &    -0.88 & -1.20 & -0.71 \\
    Min                &   -8.95 & -4.71 & -4.39 & -3.02 & -5.94 & -6.88 &  -3.96 & -6.29 &    -7.00 & -9.46 & -3.74 \\
    Max                &   11.35 &  3.78 &  4.83 &  4.49 &  2.53 &  2.75 &   3.26 &  7.94 &     6.35 & 14.53 &  5.04 \\
    \midrule
    Publication year                &   1972 &  1993 &  1993 &  2015 &  2015 &  2014 &   2014 &  2014 &     2013 & 1997 &  2013 \\
    \bottomrule
    \end{tabular}
    \label{tab:table1}
\end{table*}

Summary statistics are provided in \Cref{tab:table1}.
All factors display positive annualised average compounded returns over the considered period. Moreover, the volatility value significantly varies among factors.
In particular, Mkt-RF shows the highest value whereas the investment attitude relating factors (CMA and R-IA) display the lowest ones.
Overall, according to risk-adjusted returns, BAB and R-ROE factors turn out to be the highest-performing risk premia.
The previous observation is also supported by the Sortino ratio.\footnote{A metric to evaluate the risk-adjusted performance of a portfolio discounting for its downside standard deviation.}
We also report the release date of each factor, and note that the majority of the factors were published after 2010.

\begin{figure}[t]
    \centering
    \includegraphics[width=.35\textwidth]{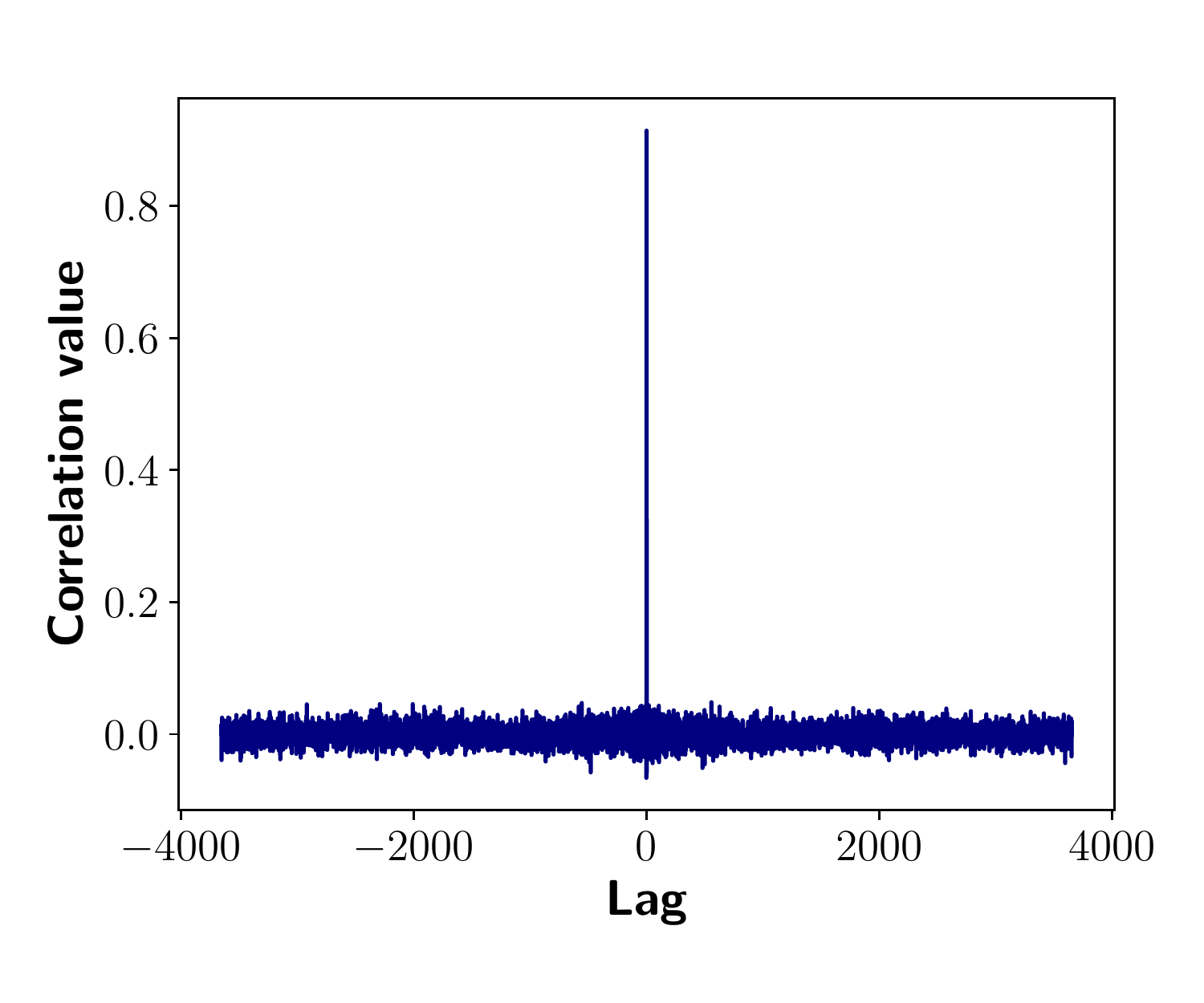}
    \vspace{-\baselineskip}
    \caption{Cross-correlation function between CMA (Conservative Minus Aggressive) and R-IA (HXZ Investment).}
    \label{fig:ccf}
\end{figure}

Finally, by inspecting the cross-correlation function (CCF), we observe significant values, especially within factors which aim to capture the same anomaly in stock returns.
Moreover, cross-correlation tends to be higher at lag $l=0$ and then to significantly drop in almost all cases.
As an example, \Cref{fig:ccf} depicts the CCF for the pair of previously-mentioned factors, CMA and R-IA.
In particular, the CCF shows a peak at lag $l=0$ of about $0.90$ and then plummets for higher order lags.
Thus, previous observations concerning the CCF behaviour suggest that there is an important ongoing associative dynamic among factors.

\subsection{Fear and business cycle indicators}
\label{sec:indicators}

In order to  relate the evolution of risk factor interactions to both stock market volatility and macroeconomic downturn, in addition to factors data, we consider two indicators defined on
the VIX Index and the yield spread, respectively.
The former index measures 30-day-ahead investors expectation of US equity market volatility.
In particular, the VIX Index is widely referred as the \emph{fear index}, since it is an indicator of market stress and financial turmoil.
The latter spread is a macroeconomic indicator which is largely used to predict recessions~\cite{estrella1996yield}.
In particular a growth in the difference between 3-month and 10-year US Treasury yield is linked to a worsening of the macroeconomic environment~\cite{estrella1991term}.
We obtain data concerning these indexes from the CBOE and FRED repositories, respectively.

Starting from these two indexes, we build \emph{fear} and \emph{business cycle}  z-scores as follows.
We first define the $\Delta$VIX \emph{historical expected shortfall} for the $k$-th inference sample period $S_k$:
$\Delta \text{VIX-}ES_{k} = \mathop{\mathbb{E}}[\mathbf{V}_k^{95}]$, defined over $\mathbf{V}_k^{95} = \{v_i \in S_k |v_i>v_k^{95}\}$,
where $v_k^{95}$ is the 95-percentile value of the percent change between VIX Index closing and opening daily values.
By computing $\Delta \text{VIX-}ES_{k}$ we quantify the extreme values of the daily volatility swing over the inference period $S_k$.
Next, we measure the extent to which such a value is unusual with respect to past observations.
Therefore, we define:
\begin{equation*}
    \text{f-zscore} = \dfrac{\Delta\text{VIX-}ES_{k}-\mu_k^{10Y}}{\sigma_k^{10Y}}
\end{equation*}
with $\mu_k^{10Y}$ and $\sigma_k^{10Y}$ being the 10-year rolling average and standard deviation of $\Delta$VIX-$ES_k$ respectively.

For the \emph{business cycle z-score}, we first evaluate the extreme values of the 3M10Y yield spread for sample $S_k$, i.e.,
$\text{3M10Y-}ES_{k} = \mathop{\mathbb{E}}[\mathbf{B}_k^{95}],$ defined over $\mathbf{B}_k^{95} = \{b_i \in S_k | b_i>b_k^{95}\}$,
where, $b_k^{95}$ is the 95-percentile value of the difference between the 3-month and 10-year US Treasury daily rates.
Thus, we define the z-score as:
\begin{equation*}
    \text{bc-zscore} = \dfrac{\Delta\text{3M10Y-}ES_{k}-\mu_k^{10Y}}{\sigma_k^{10Y}}
\end{equation*}
where $\mu_k^{10Y}$ and $\sigma_k^{10Y}$ represent the 10-year rolling average and standard deviation of $\Delta$3M10Y-$ES_k$ respectively.

\subsection{Causal inference procedure and regression model}
\label{sec:meth}
As mentioned earlier, to study the dynamic of the causal structure along time, we adopt a sliding window approach.
More precisely, we divide the overall analysis period into windows $S_k$ of length $18$ months each,
with a sliding step of 3 months, obtaining 111 inference periods $S_k$.

As described in \Cref{sec:relwork}, we apply the VAR-LiNGAM algorithm to infer the causal model.
In particular, the algorithm first fits a VAR model on the data, and then estimates on the regression residuals a linear non-Gaussian causal inference method, the DirectLiNGAM~\cite{shimizu2011directlingam}.
Other existing linear non-Gaussian approaches leverage \emph{independent component analysis} (ICA~\cite{hyvarinen1999fast}) to estimate the matrix of instantaneous causal effect $\mathbf{W}^0$. The DirectLiNGAM model was proposed to solve the potential convergence issues of ICA-based methods~\cite{himberg2004validating}.
After the fit of VAR model on data, suppose to regress the residuals associated with factor $j$ on those of factor $i$, $\forall j \in \{1,\ldots,N\} \, | \,j\neq i$.
Then, the residual $z_i$ is exogeneous to the system if it is independent of the regression residual $r_j^{i}= z_j - (cov(z_j,z_i)/var(z_i)) z_i$.
The algorithm starts with an empty causal ordering set $\mathcal{O}$ and, iteratively, appends the variable which is the most independent of its residual.
The procedure stops when $N-1$ insertions have been made. 

In each sample period, we apply the model by selecting the number of lags according to the BIC criterion~\cite{schwarz1978estimating}: the resulting maximum lag $L$ is equal to 1 for every sample.
Subsequently, we validate the estimated causal coefficients by running a permutation test, i.e., resampling with replacement, with a significance level equal to $95\%$.
In addition, the total number of permuted samples per period is 5000, the length of the generated samples equals that of the inference period (18 months) and we do not apply any thresholding to the resulting significant coefficients. 
Since we are interested in comparing the information provided by causal inference with that coming from correlation analysis, the same methodology is used to estimate correlation networks, by replacing the estimation of \Cref{eq:SVAR} with the Pearson correlation coefficient.

Once we retrieve both causal and correlation network structures, we analyse their temporal evolution by means of a regression analysis.
We set as dependent variable $d$ the number of network edges and as covariates the following three variables: time (measured in days), f-zscore, and bc-zscore.
Moreover, since $d \in \mathbb{N}_0$, we employ a \emph{Poisson log-linear model} specified by the following GLM~\cite{agresti2018introduction}:
\begin{equation}\label{eq:logmod}
    log(d)=\beta_{0} + \sum_{i=1}^{3}{\beta_i \cdot f_i},
\end{equation}
\noindent
from which we have $d=e^{\beta_0} \cdot \prod_{i=1}^{3}e^{\beta_i \cdot f_i}$, where $f_i$ are the regressors mentioned before.
 Therefore, according to \Cref{eq:logmod}, a unit increase in the independent variable $f_i$ is associated with a multiplicative effect $e^{\beta_i}$ on $d$.
 As a consequence, if $\beta_i=0$, the growth of $f_i$ does not affect that of $d$.
 Furthermore, if $\beta_i>0$ then $d$ increases as $f_i$ grows, and conversely if $\beta_i<0$ it decreases. 
\section{Results}
\label{sec:results}

\begin{figure*}[t]
    \centering
    \includegraphics[width=0.8\textwidth]{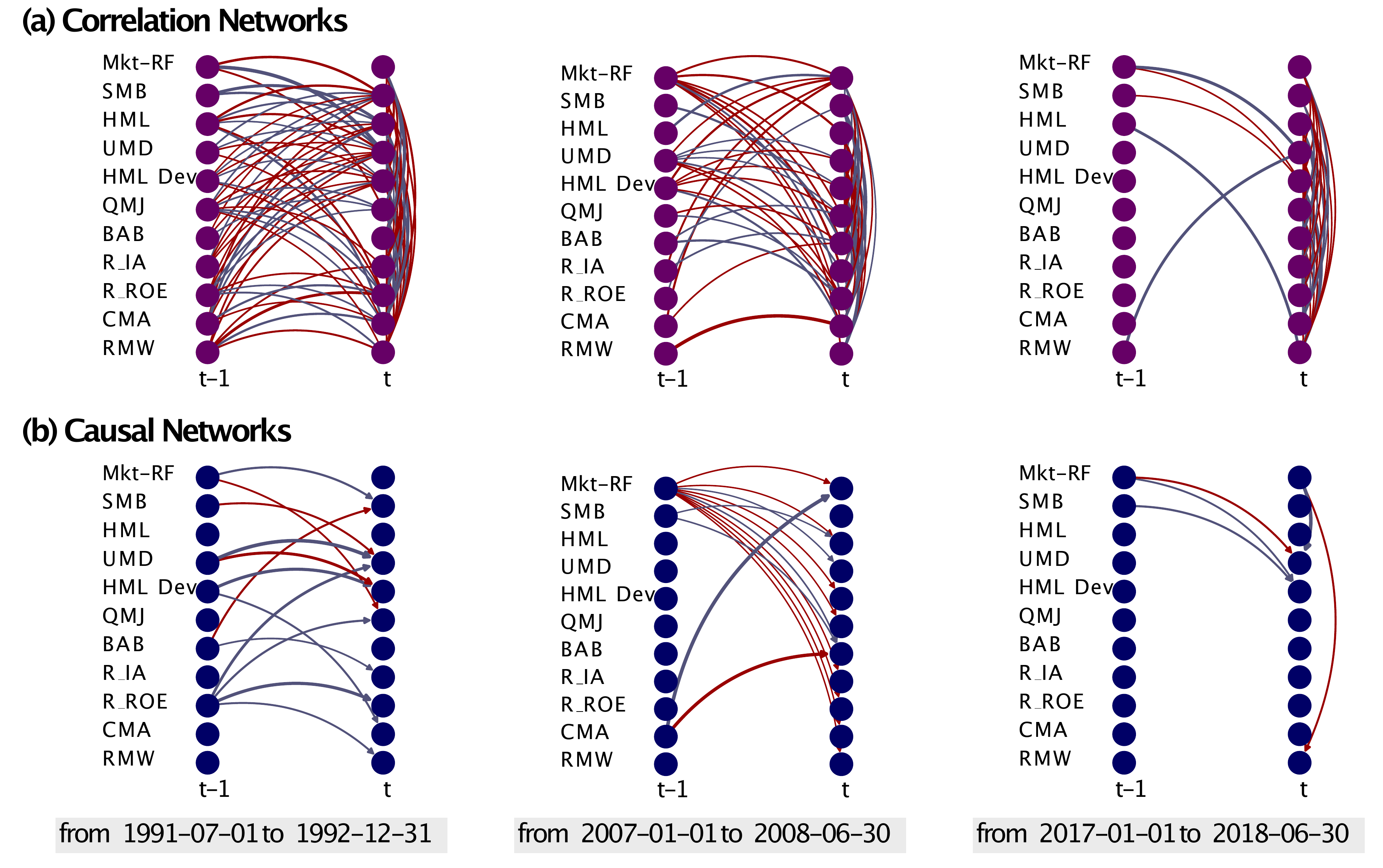}
    \caption{Correlation (a) and causal (b) structures inferred over three different periods. Factors are sorted in ascending order from the oldest to the newest. Grey edges represent positive weights, red edges negative ones.}
    \label{fig:Networks}
\end{figure*}

In this section we provide the results of the analysis. 
The inferred networks for causal and correlation structures are shown in \Cref{fig:Networks}.
For readability, only three different inference samples are shown: 
(i) before the publication of the majority of the factor models; 
(ii) during the Global Financial Crisis (GFC), located between 2007 and 2009; 
(iii) after the publication of all factor models. 

The networks are constituted by 22 nodes, split in accordance to time ordering.
The factors are sorted vertically according to their publication date, from the oldest to the newest.
In addition, edges associated with a positive weight are shown in grey, while those with a negative one are given in red.
Thicker lines indicate a higher weight of the edge, and thus a stronger association of the factors.
Finally, in order to make figures easier to read, throughout the paper results relating to correlation are shown in purple and those relating to causation in blue.

We notice a significant variability in both correlation and causal network structures across the periods.
As time goes by, the number of edges decreases and the networks become sparser.
Such phenomenon is more pronounced for inter-layer relationships. 
\Cref{fig:Networks} highlights the key role of market factor, which impacts 9 factors out of 11 during the GFC. 


\begin{figure}[t]
    \centering
     \includegraphics[width=0.35\textwidth]{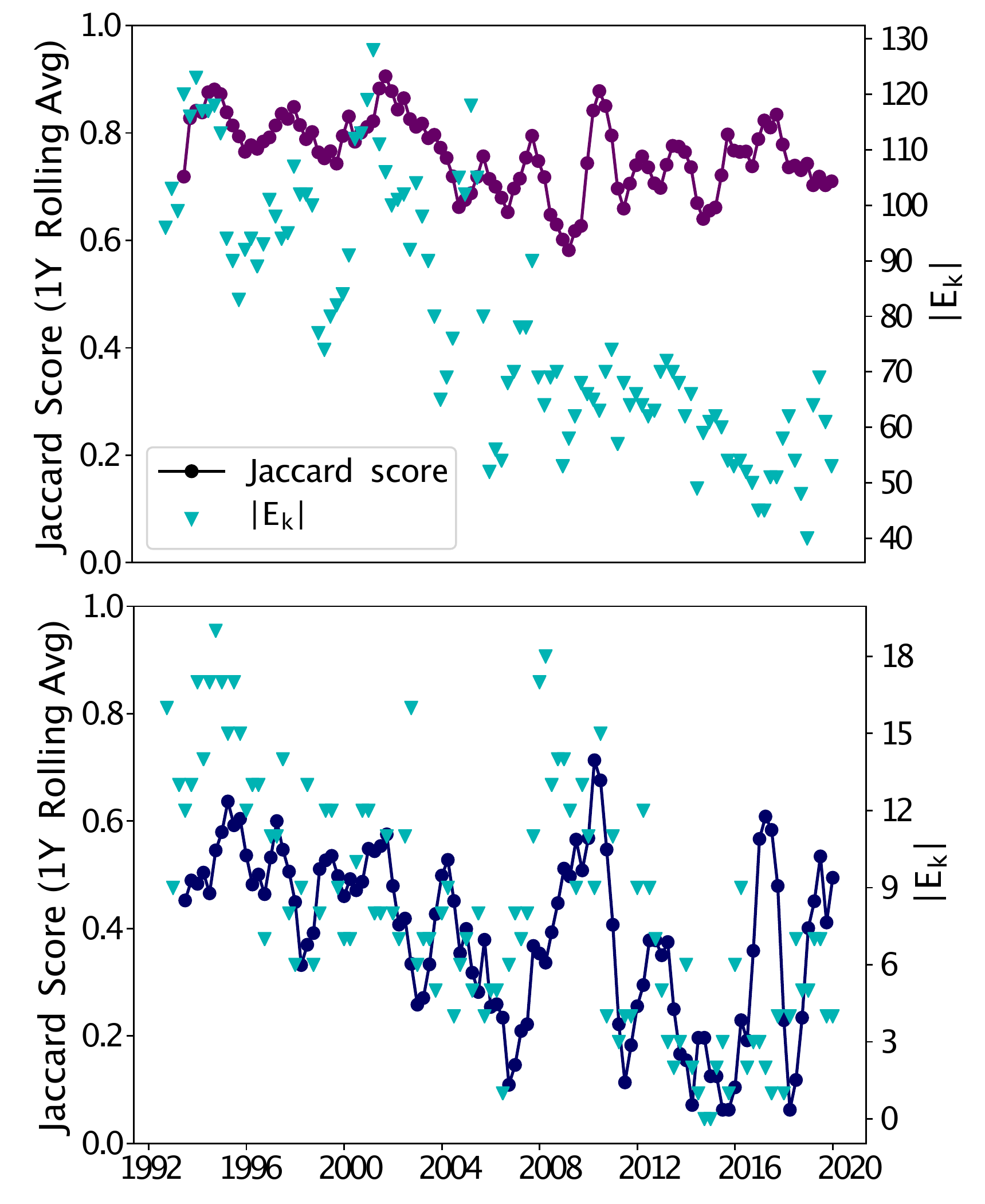}
      \caption{Behaviour over time of 1-year rolling average Jaccard score between consecutive correlation networks (top) and causal networks (bottom). The right axis shows the number of network edges.}
    \label{fig:nonstationarity}
\end{figure}
    
As far as the stability of the inferred networks is concerned, \Cref{fig:nonstationarity} shows the evolution over time of Jaccard score between two networks estimated on adjacent periods.
Such a metric is used in order to quantify the matching ratio between the edge sets $\mathcal{E}_{k-1}$ and $\mathcal{E}_{k}$ of two consecutive structures. 
In particular, it is defined as:
\begin{equation*}
    Jacc(\mathcal{E}_{k-1},\mathcal{E}_{k}) = \dfrac{|\mathcal{E}_{k-1} \cap \mathcal{E}_{k}|}{|\mathcal{E}_{k-1} \cup \mathcal{E}_{k}|}
\end{equation*}
\noindent
and it holds $Jacc(\mathcal{E}_{k-1},\mathcal{E}_{k}) \in [0,1]$. 
In order to smooth the score, we apply a 1-year rolling average filter: as we use a step size of 3 months, the average of the last four values is shown.
The Jaccard score is already normalized and is not affected by the size of the sets, however when the numbers are very small it could be misleading.
To better interpret the plot, the cardinality of the edge set $|\mathcal{E}_{k}|$ is given on the right y-axis of each chart. 
When the correlation networks are considered, the aforementioned score is more stable over time and is much higher than the one returned by causal structures.
In both cases, we observe the presence of a sparsifying trend over time.
Interestingly, for causal networks, this trend breaks down during the GFC, and the Jaccard score sharply increases.


\begin{table}[t]
    \centering
    \small
    \footnotesize
    \caption{Results of the GLM regression  of the number of edges in \emph{correlation} networks. Bold coefficients are statistically significant at 99\% level.}
    \begin{tabular}{llrrr}
    \toprule
    {Relation type} & {variable} & {coef.} & {std. err.} & {p-value} \\
    \midrule
    \multirow{4}{*}{Overall}
        & \textbf{intercept}      &       \textbf{4.7419}  &        0.024     &         0.000        \\
        & \textbf{time}             &      \textbf{-7.054e-05}  &      4.3e-06     &         0.000        \\
        & \textbf{f-zscore}       &       \textbf{-0.0402}  &        0.009     &         0.000        \\
        & bc-zscore    &       -0.0039  &        0.010     &         0.687        \\
    \cline{1-5}
    \multirow{4}{*}{Instantaneous} 
        & \textbf{intercept}      &     \textbf{3.7993}  &        0.035     &         0.000             \\
        & time             &       4.07e-06  &     5.72e-06     &         0.477        \\
        & \textbf{f-zscore}       &       \textbf{-0.0340}  &        0.012     &         0.004        \\
        & bc-zscore    &      -0.0038  &        0.012     &         0.761        \\
    \cline{1-5}
    \multirow{4}{*}{Lagged}
        & \textbf{intercept}      &       \textbf{4.3698}  &        0.034     &         0.000        \\
        & \textbf{time}             &      \textbf{-0.0002}  &     6.76e-06     &         0.000       \\
        & \textbf{f-zscore}       &      \textbf{-0.0718}  &        0.014     &         0.000        \\
        & bc-zscore    &       -0.0229  &        0.015     &         0.131        \\
    \bottomrule
    \end{tabular}
    \label{tab:correl}
\end{table}

We further analyze the temporal trend of network density by means of a regression analysis.
In particular, we relate the number of edges to time, fear, and business cycle indicators through the estimation of \Cref{eq:logmod}.
Results related to correlation structures are reported in \Cref{tab:correl}.

Considering the overall relations, both time and f-zscore are statistically significant at $99\%$ level and are related to the sparsification of the network.
It is worth noticing that time is measured in days, and thus the magnitude of the estimated coefficient is expected to be very small.
Relating the value of the coefficient to the length of the considered time window, we obtain that every 18 months the number of edges in the correlation structure is reduced by approximately $4\%$.
With regard to investors future expectation, a growth of one standard deviation in f-zscore is associated with a reduction of almost $4\%$ in the number of edges as well.
On the contrary, the bc-zscore is never statistically significant.
Thus, the results suggest that the correlation structure does not show a link with changes in macroeconomic conditions.

\begin{figure}[t]
    \centering
    \captionsetup[subfigure]{aboveskip=-5pt,belowskip=-5pt}
    \begin{subfigure}[b]{.235\textwidth}
        \centering
        \includegraphics[width=\textwidth]{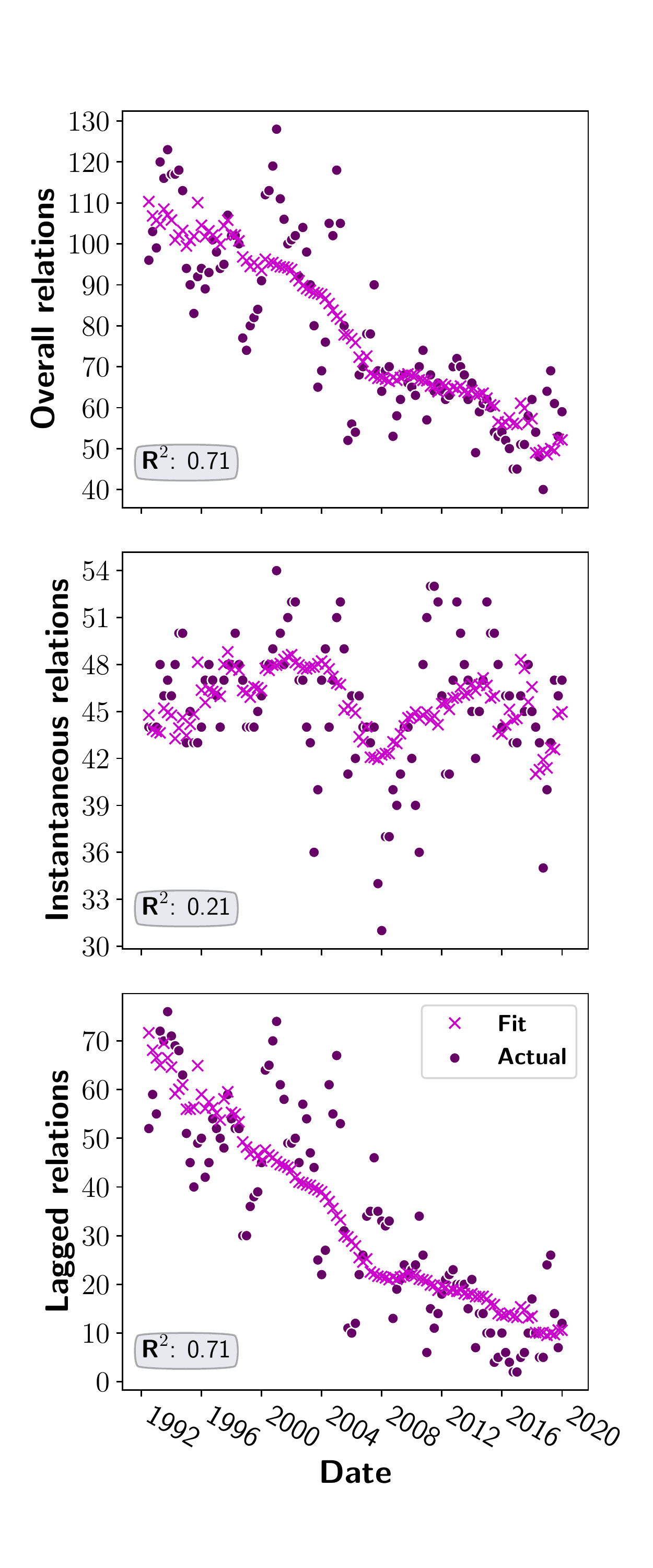}
        \caption{Correlation Networks}
        \label{fig:regr_corr}
    \end{subfigure}
    \hfill
    \begin{subfigure}[b]{.235\textwidth}
        \centering
        \includegraphics[width=\textwidth]{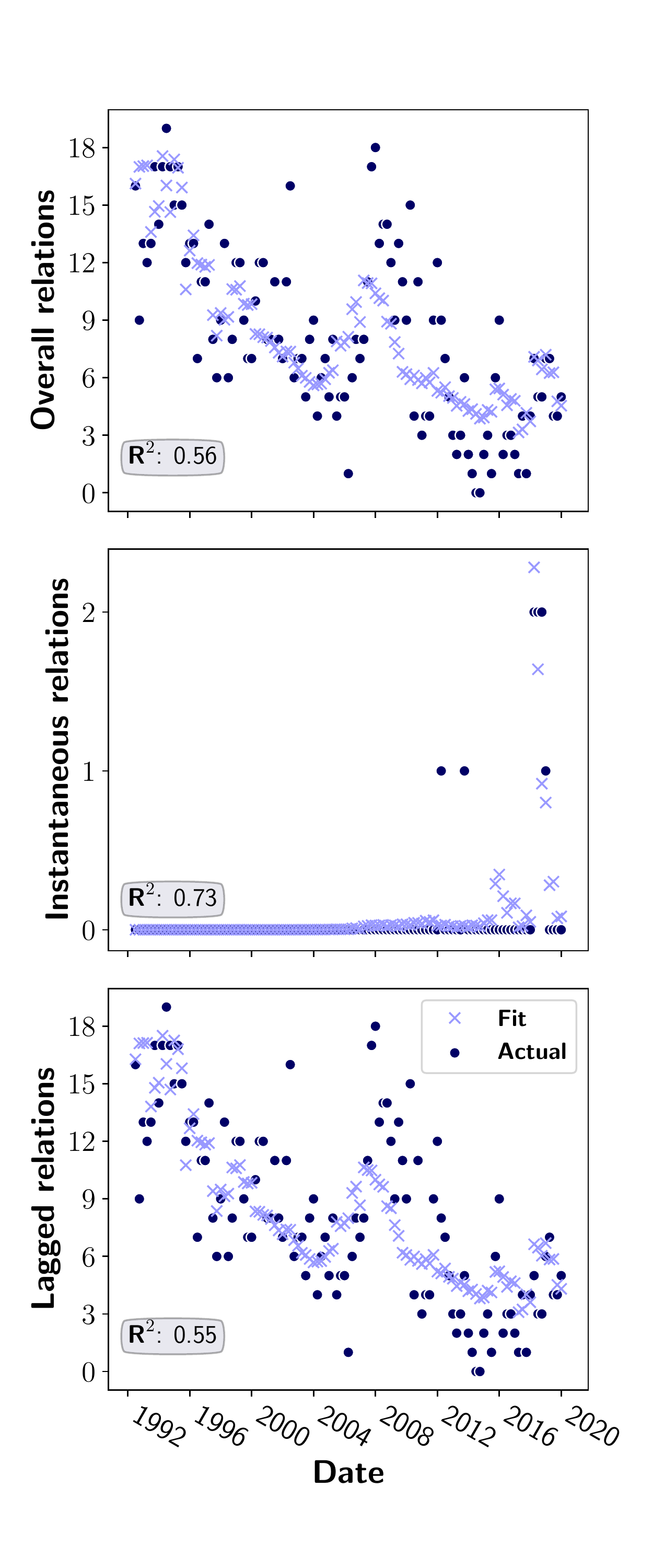}
        \caption{Causal Networks}
        \label{fig:regr_causal}
    \end{subfigure}
    \vspace{-\baselineskip}
    \caption{Evolution of the estimated number of significant edges in both (a) correlation and (b) causal networks.}
\label{fig:regressions}
\end{figure}

In addition to overall relations, we analyse instantaneous and lagged interactions separately.
In particular, while the f-zscore remains significant in both cases, time is only relevant for lagged relations.
Regression results and \Cref{fig:regr_corr} (in which we provide the fit of the observational data) show that, even though we observe a slight decrease during stress periods, the level of instantaneous association is pretty stable over time.
Thus, the overall statistical significance of time is due to the disappearance of inter-layer links.

\begin{table}[t]
    \centering
    \small
    \footnotesize
    \caption{Results of the GLM regression  of the number of edges in \emph{causal} networks. Bold coefficients are statistically significant at 99\% level.}
    \begin{tabular}{llrrr}
    \toprule
    {Relation type} & {variable} & {coef.} & {std. err.} & {p-value} \\
    \midrule
    \multirow{4}{*}{{Overall}}
        & \textbf{intercept}      &       \textbf{2.8670}  &        0.067     &         0.000        \\
        & \textbf{time}             &      \textbf{-0.0002}  &     1.29e-05     &         0.000        \\
        & \textbf{f-zscore}       &       \textbf{0.1732}  &        0.027     &         0.000        \\
        & {bc-zscore}    &       0.0782  &        0.030     &         0.010        \\
    \cline{1-5}
    \multirow{4}{*}{{Instantaneous}} 
        & {intercept}      &     -10.5020  &        4.188     &         0.012             \\
        & {time}             &       0.0008  &        0.000     &         0.102        \\
        & \textbf{f-zscore}       &       \textbf{1.0139}  &        0.357     &         0.005        \\
        & {bc-zscore}    &      -0.4617  &        0.415     &         0.266        \\
    \cline{1-5}
    \multirow{4}{*}{{Lagged}}
        & \textbf{intercept}      &       \textbf{2.8790}  &        0.067     &         0.000        \\
        & \textbf{time}             &      \textbf{-0.0002}  &     1.29e-05     &         0.000       \\
        & \textbf{f-zscore}       &      \textbf{0.1640}  &        0.027     &         0.000        \\
        & {bc-zscore}    &       0.0727  &        0.030     &         0.017        \\
    \bottomrule
    \end{tabular}
    \label{tab:causal}
\end{table}

As far as causal networks are concerned, \Cref{tab:causal} provides the results of the regression analysis.
By focusing on the overall relations, we see that both time and fear index are statistically significant at $99\%$ level. 
Here, the network sparsification is faster, i.e., every 18 months the total number of connections decreases by about $11\%$.
On the contrary, an increase of one standard deviation in the fear index is associated with a growth of almost $19\%$ in the number of edges. 
This finding explains the break in the sparsifying trend during periods of financial stress shown in \Cref{fig:nonstationarity}.
Regarding the business cycle indicator, it is significant at $95\%$ level, with the sign of the regression coefficient in accordance with that shown by the fear index. 
Therefore, the analysis of causal structures shows that a worsening in the macroeconomic environment is also linked to an increase of causal relations among factors. 
However, the strength of the effect is much smaller than that of the fear index. 

By analysing instantaneous relations apart from lagged ones, we notice that the statistical significance of the fear index is preserved, whereas time is only relevant for lagged connections.
The estimated coefficient indicates that the network of lagged interactions thins out with the same intensity as that of all the connections.
Such a phenomenon is illustrated in \Cref{fig:regr_causal}: the trend of the causal structure mainly consists of lagged causal interactions while instantaneous effects only appear in recent years.
Interestingly, the temporal trend of network sparsification is similar to the one shown above for lagged connections in correlation networks.

\begin{figure}[t]
    \centering
    \includegraphics[width=.35\textwidth]{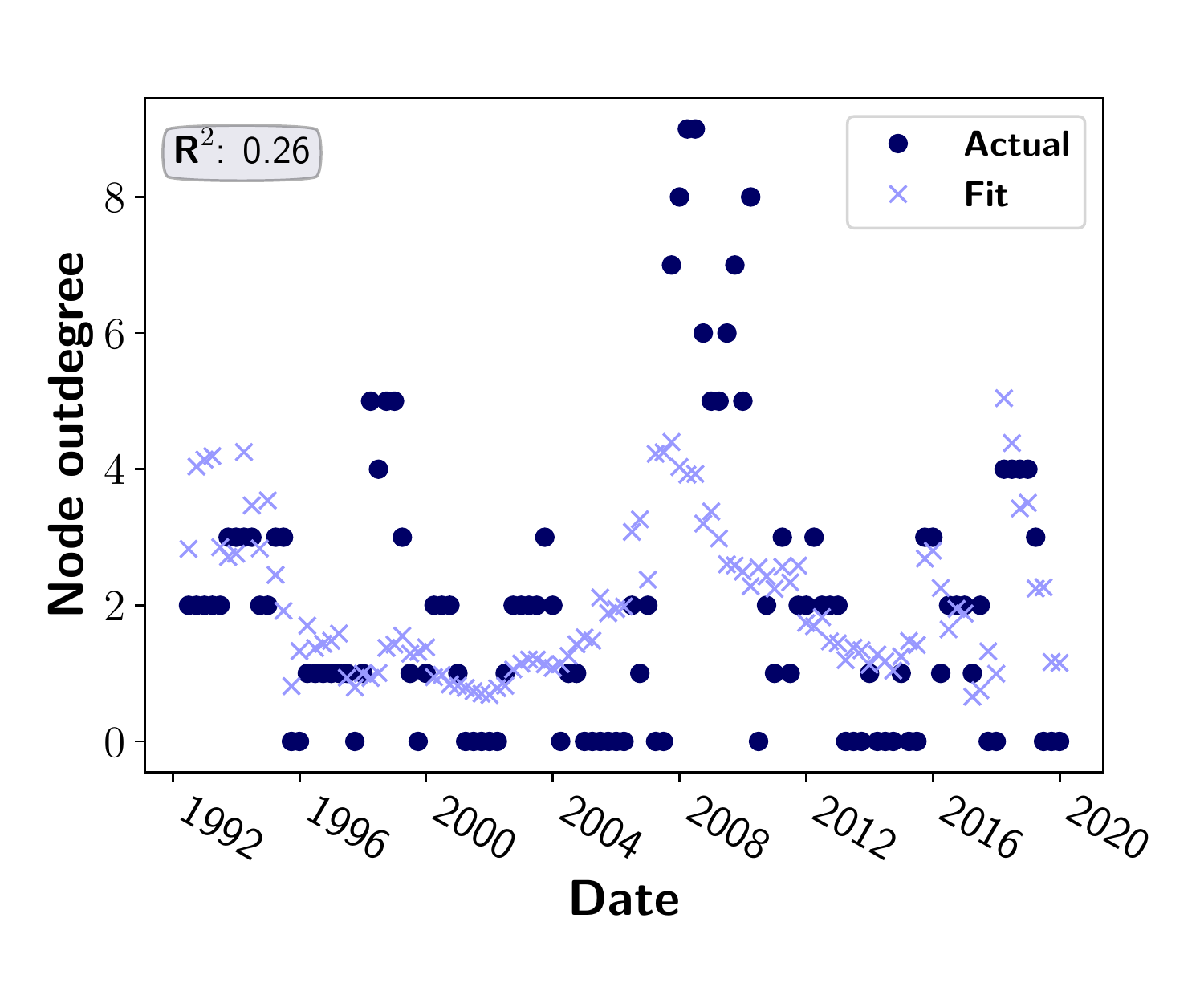}
    \vspace{-\baselineskip}
    \caption{Evolution over time (and GLM fit) of the out-degree of the market risk factor in causal networks.}
    \label{fig:mkt_causal}
\end{figure}

Finally, we focus on the role of the market risk factor node within the network of causal relations among risk factors. 
As shown in \Cref{fig:mkt_causal}, the out-degree of the corresponding node remains in the range of 2 most of the time, then dramatically increases during periods of crisis, as observed during the GFC and more recently during 2018. 
In the latter example the stock market suffered heavy losses: 
first a volatility shock occurred in early February;
subsequently the market plummeted in the last quarter due to both US-China trade war and the slowdown in economic growth.  

\begin{table}[t]
    \centering
    \small
    \caption{Results of the GLM regression  of the out-degree of the market risk factor in causal networks.}
    \resizebox{.6\columnwidth}{!}{%
    \begin{tabular}{lrrr}
    \toprule
    {variable} & {coef.} & {std. err.} & {p-value} \\
    \midrule
    \textbf{intercept}                   &       \textbf{1.0963}  &        0.144     &         0.000        \\
    \textbf{time}                    &      \textbf{-0.0001}  &     2.57e-05     &         0.000        \\
    \textbf{f-zscore} &       \textbf{0.4382}  &        0.058     &         0.000        \\
    {bc-zscore}          &       -0.0790  &        0.063     &         0.210        \\
    \bottomrule
    \end{tabular}%
    }
    \label{tab:causal_mkt}
\end{table}

In order to investigate the evolution over time of market node out-degree, \Cref{tab:causal_mkt} reports the results of a regression analysis.
We set as a dependent variable the aforementioned out-degree and as covariates time, f-zscore, and bc-zscore. 
We use again the Poisson log-linear model described by \Cref{eq:logmod}.
The association with the fear indicator is statistically significant at $99\%$ level:
an increase of one standard deviation in the latter is linked to a strong growth of almost $55\%$ in the node out-degree.
Furthermore, time turns out to be a significant feature, which is consistent with the already underlined presence of the sparsifying trend in causal relations.

\section{Discussion}
\label{sec:discussion}
The results shown in the present work highlight the continuously-changing nature of risk factors interactions.
Taking such behaviours into account is of paramount importance for implementing an effective risk management process in multi-factor investing:
given the high non-stationarity of the system at hand, it is fundamental to develop robust causal structure learning models for dealing with small samples, and to analyze the causal interactions of risk premia at a finer grain.

\spara{Sparsification.}
As far as the changing in factors relationships over time is concerned, our results support, from a causal perspective, the evidence of factor redundancy that have been provided by recent findings~\cite{feng2020taming,HARVEY2021}.
The analysis of the inferred causal networks suggests that, during past years the factorial system was driven by one-lag causal interactions whereas, more recently, instantaneous relationships among those risk factors have appeared.
Therefore, a proper factor causal model needs to take into account instantaneous effects as well.
In addition, both causation and correlation analyses support the presence of a sparsifying trend in inter-layer connections.
In particular, we hypothesize that the loss of memory of the system may be due to an increase in the sophistication of the market participants, who are able to react faster to external stimuli~\cite{brogaard2014high, chordia2018rent}.

\spara{Factor unveiling.}
To better characterize the evolution over time of the system under consideration, we checked whether the phenomenon of unveiling a factorial model would affect the causal structure.
The rationale behind such a test is that, after a factor becomes known, investors start betting on it, and then factor relations might be altered.
However, we did not find any statistically significant evidence for the link between unveiling a factor and a change in the causal structure around it.
In order to identify the drivers of the aforementioned sparsifying trend, it might be useful to analyze the system at hand by using higher frequency data and, in addition, to inspect possible links with the recent commodification of factorial strategies within the US market. 

\spara{Financial and economic stress.}
With regard to the behavior of the factorial system during stress periods, inspecting both causation and correlation results provides a richer view of the underlying dynamics.
Our findings show that, during financial turmoil, the level of instantaneous association among factors slightly decreases, and the factorial system becomes driven by the market factor.
As a consequence, the causal structure becomes denser and more stable due to the increase of the out-degree of the market node, i.e., the influence that the market has on all the other factors.
This result is of paramount importance for investors, since it shows that, during stress periods, the exposure to different factors reverts to a simple exposure to market risk.
Our finding echoes recent ones which provide evidence for the market factor being by far the dominant one~\cite{giglio2017inference, HARVEY2021}.
Furthermore, the reported relationship between the change in the VIX Index and both correlation and causal structures reinforces existing evidence in financial literature concerning the link between the VIX Index and factors returns~\cite{durand2011fear}.

Finally, we have analyzed the relation between the change in macroeconomic conditions, as measured by the spread of the yield curve, and the evolution of the factorial system. Also in this case we can appreciate the benefit of looking at causation.
According to correlation analysis alone, the business cycle indicator is not associated with the dynamics of the analyzed system.
Conversely, the regression analysis which relates the business cycle indicator to the causal structure of the factorial system displays a statistically significant relationship with a $95\%$ confidence level.
Indeed, similarly to the results of the volatility analysis, the worsening of macroeconomic environment is associated with a growth in the number of network arcs.
Therefore, looking at the causation allows to better inspect the evolution of the system during negative economic phases.

\spara{Future work.}
The results in this paper contribute to make some progress in understanding the relationships among risk factors.
However, several questions remain open.
As an example, it would be interesting to study the interactions among risk factors belonging to different equity markets.
Moreover, our analysis concerns only the equity asset class.
Thus, enlarging the considered set of factors, by including risk premia concerning other asset classes as well, could help in taking into consideration also inter-asset class dynamics.
Finally, by construction, causal networks enable studying the response of the system at hand under interventions.
Therefore, it would be interesting to exploit the attained results to setup a suitable stress testing procedure for multi-factor portfolios.

\begin{acks}
The authors acknowledge the support from Intesa Sanpaolo Innovation Center.
The funder had no role in study design, data collection and analysis, decision to publish, or preparation of the manuscript.
\end{acks}

\bibliographystyle{ACM-Reference-Format}
\bibliography{bibliography}


\begin{thebibliography}{48}


\ifx \showCODEN    \undefined \def \showCODEN     #1{\unskip}     \fi
\ifx \showDOI      \undefined \def \showDOI       #1{#1}\fi
\ifx \showISBNx    \undefined \def \showISBNx     #1{\unskip}     \fi
\ifx \showISBNxiii \undefined \def \showISBNxiii  #1{\unskip}     \fi
\ifx \showISSN     \undefined \def \showISSN      #1{\unskip}     \fi
\ifx \showLCCN     \undefined \def \showLCCN      #1{\unskip}     \fi
\ifx \shownote     \undefined \def \shownote      #1{#1}          \fi
\ifx \showarticletitle \undefined \def \showarticletitle #1{#1}   \fi
\ifx \showURL      \undefined \def \showURL       {\relax}        \fi
\providecommand\bibfield[2]{#2}
\providecommand\bibinfo[2]{#2}
\providecommand\natexlab[1]{#1}
\providecommand\showeprint[2][]{arXiv:#2}

\bibitem[\protect\citeauthoryear{Agresti}{Agresti}{2018}]%
        {agresti2018introduction}
\bibfield{author}{\bibinfo{person}{Alan Agresti}.}
  \bibinfo{year}{2018}\natexlab{}.
\newblock \bibinfo{booktitle}{\emph{An introduction to categorical data
  analysis}}.
\newblock \bibinfo{publisher}{John Wiley \& Sons}. 74--90 pages.
\newblock


\bibitem[\protect\citeauthoryear{Asness and Frazzini}{Asness and
  Frazzini}{2013}]%
        {asness2013devil}
\bibfield{author}{\bibinfo{person}{Clifford Asness} {and}
  \bibinfo{person}{Andrea Frazzini}.} \bibinfo{year}{2013}\natexlab{}.
\newblock \showarticletitle{The devil in HML’s details}.
\newblock \bibinfo{journal}{\emph{The Journal of Portfolio Management}}
  \bibinfo{volume}{39}, \bibinfo{number}{4} (\bibinfo{year}{2013}),
  \bibinfo{pages}{49--68}.
\newblock


\bibitem[\protect\citeauthoryear{Asness, Frazzini, and Pedersen}{Asness
  et~al\mbox{.}}{2019}]%
        {asness2019quality}
\bibfield{author}{\bibinfo{person}{Clifford~S Asness}, \bibinfo{person}{Andrea
  Frazzini}, {and} \bibinfo{person}{Lasse~Heje Pedersen}.}
  \bibinfo{year}{2019}\natexlab{}.
\newblock \showarticletitle{Quality minus junk}.
\newblock \bibinfo{journal}{\emph{Review of Accounting Studies}}
  \bibinfo{volume}{24}, \bibinfo{number}{1} (\bibinfo{year}{2019}),
  \bibinfo{pages}{34--112}.
\newblock


\bibitem[\protect\citeauthoryear{Bardoscia, Barucca, Battiston, Caccioli,
  Cimini, Garlaschelli, Saracco, Squartini, and Caldarelli}{Bardoscia
  et~al\mbox{.}}{2021}]%
        {bardoscia2021physics}
\bibfield{author}{\bibinfo{person}{Marco Bardoscia}, \bibinfo{person}{Paolo
  Barucca}, \bibinfo{person}{Stefano Battiston}, \bibinfo{person}{Fabio
  Caccioli}, \bibinfo{person}{Giulio Cimini}, \bibinfo{person}{Diego
  Garlaschelli}, \bibinfo{person}{Fabio Saracco}, \bibinfo{person}{Tiziano
  Squartini}, {and} \bibinfo{person}{Guido Caldarelli}.}
  \bibinfo{year}{2021}\natexlab{}.
\newblock \showarticletitle{The Physics of Financial Networks}.
\newblock \bibinfo{journal}{\emph{arXiv preprint arXiv:2103.05623}}
  (\bibinfo{year}{2021}).
\newblock


\bibitem[\protect\citeauthoryear{Billio, Getmansky, Lo, and Pelizzon}{Billio
  et~al\mbox{.}}{2012}]%
        {billio2012econometric}
\bibfield{author}{\bibinfo{person}{Monica Billio}, \bibinfo{person}{Mila
  Getmansky}, \bibinfo{person}{Andrew~W Lo}, {and} \bibinfo{person}{Loriana
  Pelizzon}.} \bibinfo{year}{2012}\natexlab{}.
\newblock \showarticletitle{Econometric measures of connectedness and systemic
  risk in the finance and insurance sectors}.
\newblock \bibinfo{journal}{\emph{Journal of financial economics}}
  \bibinfo{volume}{104}, \bibinfo{number}{3} (\bibinfo{year}{2012}),
  \bibinfo{pages}{535--559}.
\newblock


\bibitem[\protect\citeauthoryear{Bollerslev}{Bollerslev}{1986}]%
        {bollerslev1986generalized}
\bibfield{author}{\bibinfo{person}{Tim Bollerslev}.}
  \bibinfo{year}{1986}\natexlab{}.
\newblock \showarticletitle{Generalized autoregressive conditional
  heteroskedasticity}.
\newblock \bibinfo{journal}{\emph{Journal of econometrics}}
  \bibinfo{volume}{31}, \bibinfo{number}{3} (\bibinfo{year}{1986}),
  \bibinfo{pages}{307--327}.
\newblock


\bibitem[\protect\citeauthoryear{Brogaard, Hendershott, and Riordan}{Brogaard
  et~al\mbox{.}}{2014}]%
        {brogaard2014high}
\bibfield{author}{\bibinfo{person}{Jonathan Brogaard},
  \bibinfo{person}{Terrence Hendershott}, {and} \bibinfo{person}{Ryan
  Riordan}.} \bibinfo{year}{2014}\natexlab{}.
\newblock \showarticletitle{High-frequency trading and price discovery}.
\newblock \bibinfo{journal}{\emph{The Review of Financial Studies}}
  \bibinfo{volume}{27}, \bibinfo{number}{8} (\bibinfo{year}{2014}),
  \bibinfo{pages}{2267--2306}.
\newblock


\bibitem[\protect\citeauthoryear{B{\"u}hlmann, Peters, Ernest,
  et~al\mbox{.}}{B{\"u}hlmann et~al\mbox{.}}{2014}]%
        {buhlmann2014cam}
\bibfield{author}{\bibinfo{person}{Peter B{\"u}hlmann}, \bibinfo{person}{Jonas
  Peters}, \bibinfo{person}{Jan Ernest}, {et~al\mbox{.}}}
  \bibinfo{year}{2014}\natexlab{}.
\newblock \showarticletitle{CAM: Causal additive models, high-dimensional order
  search and penalized regression}.
\newblock \bibinfo{journal}{\emph{Annals of statistics}} \bibinfo{volume}{42},
  \bibinfo{number}{6} (\bibinfo{year}{2014}), \bibinfo{pages}{2526--2556}.
\newblock


\bibitem[\protect\citeauthoryear{Carhart}{Carhart}{1997}]%
        {carhart1997persistence}
\bibfield{author}{\bibinfo{person}{Mark~M Carhart}.}
  \bibinfo{year}{1997}\natexlab{}.
\newblock \showarticletitle{On persistence in mutual fund performance}.
\newblock \bibinfo{journal}{\emph{The Journal of finance}}
  \bibinfo{volume}{52}, \bibinfo{number}{1} (\bibinfo{year}{1997}),
  \bibinfo{pages}{57--82}.
\newblock


\bibitem[\protect\citeauthoryear{Chickering}{Chickering}{2002}]%
        {chickering2002}
\bibfield{author}{\bibinfo{person}{David~Maxwell Chickering}.}
  \bibinfo{year}{2002}\natexlab{}.
\newblock \showarticletitle{Optimal structure identification with greedy
  search}.
\newblock \bibinfo{journal}{\emph{Journal of machine learning research}}
  \bibinfo{volume}{3}, \bibinfo{number}{Nov} (\bibinfo{year}{2002}),
  \bibinfo{pages}{507--554}.
\newblock


\bibitem[\protect\citeauthoryear{Chordia, Green, and Kottimukkalur}{Chordia
  et~al\mbox{.}}{2018}]%
        {chordia2018rent}
\bibfield{author}{\bibinfo{person}{Tarun Chordia}, \bibinfo{person}{T~Clifton
  Green}, {and} \bibinfo{person}{Badrinath Kottimukkalur}.}
  \bibinfo{year}{2018}\natexlab{}.
\newblock \showarticletitle{Rent seeking by low-latency traders: Evidence from
  trading on macroeconomic announcements}.
\newblock \bibinfo{journal}{\emph{The Review of Financial Studies}}
  \bibinfo{volume}{31}, \bibinfo{number}{12} (\bibinfo{year}{2018}),
  \bibinfo{pages}{4650--4687}.
\newblock


\bibitem[\protect\citeauthoryear{Cochrane}{Cochrane}{2009}]%
        {cochrane2009asset}
\bibfield{author}{\bibinfo{person}{John~H Cochrane}.}
  \bibinfo{year}{2009}\natexlab{}.
\newblock \showarticletitle{The Cross-section: CAPM and Multifactor Models}.
\newblock In \bibinfo{booktitle}{\emph{Asset pricing (Revised edition)}}.
  \bibinfo{publisher}{Princeton university press}, Chapter~20,
  \bibinfo{pages}{435--449}.
\newblock


\bibitem[\protect\citeauthoryear{Cochrane}{Cochrane}{2011}]%
        {cochrane2011presidential}
\bibfield{author}{\bibinfo{person}{John~H Cochrane}.}
  \bibinfo{year}{2011}\natexlab{}.
\newblock \showarticletitle{Presidential address: Discount rates}.
\newblock \bibinfo{journal}{\emph{The Journal of finance}}
  \bibinfo{volume}{66}, \bibinfo{number}{4} (\bibinfo{year}{2011}),
  \bibinfo{pages}{1047--1108}.
\newblock


\bibitem[\protect\citeauthoryear{Durand, Lim, and Zumwalt}{Durand
  et~al\mbox{.}}{2011}]%
        {durand2011fear}
\bibfield{author}{\bibinfo{person}{Robert~B Durand}, \bibinfo{person}{Dominic
  Lim}, {and} \bibinfo{person}{J~Kenton Zumwalt}.}
  \bibinfo{year}{2011}\natexlab{}.
\newblock \showarticletitle{Fear and the Fama-French factors}.
\newblock \bibinfo{journal}{\emph{Financial Management}} \bibinfo{volume}{40},
  \bibinfo{number}{2} (\bibinfo{year}{2011}), \bibinfo{pages}{409--426}.
\newblock


\bibitem[\protect\citeauthoryear{Estrella and Hardouvelis}{Estrella and
  Hardouvelis}{1991}]%
        {estrella1991term}
\bibfield{author}{\bibinfo{person}{Arturo Estrella} {and}
  \bibinfo{person}{Gikas~A Hardouvelis}.} \bibinfo{year}{1991}\natexlab{}.
\newblock \showarticletitle{The term structure as a predictor of real economic
  activity}.
\newblock \bibinfo{journal}{\emph{The journal of Finance}}
  \bibinfo{volume}{46}, \bibinfo{number}{2} (\bibinfo{year}{1991}),
  \bibinfo{pages}{555--576}.
\newblock


\bibitem[\protect\citeauthoryear{Estrella and Mishkin}{Estrella and
  Mishkin}{1996}]%
        {estrella1996yield}
\bibfield{author}{\bibinfo{person}{Arturo Estrella} {and}
  \bibinfo{person}{Frederic~S Mishkin}.} \bibinfo{year}{1996}\natexlab{}.
\newblock \showarticletitle{The yield curve as a predictor of US recessions}.
\newblock \bibinfo{journal}{\emph{Current issues in economics and finance}}
  \bibinfo{volume}{2}, \bibinfo{number}{7} (\bibinfo{year}{1996}).
\newblock


\bibitem[\protect\citeauthoryear{Fama and French}{Fama and French}{1993}]%
        {fama1993common}
\bibfield{author}{\bibinfo{person}{Eugene~F Fama} {and}
  \bibinfo{person}{Kenneth~R French}.} \bibinfo{year}{1993}\natexlab{}.
\newblock \showarticletitle{Common risk factors in the returns on stocks and
  bonds}.
\newblock \bibinfo{journal}{\emph{Journal of financial economics}}
  \bibinfo{volume}{33}, \bibinfo{number}{1} (\bibinfo{year}{1993}),
  \bibinfo{pages}{3--56}.
\newblock


\bibitem[\protect\citeauthoryear{Fama and French}{Fama and French}{2015}]%
        {fama2015five}
\bibfield{author}{\bibinfo{person}{Eugene~F Fama} {and}
  \bibinfo{person}{Kenneth~R French}.} \bibinfo{year}{2015}\natexlab{}.
\newblock \showarticletitle{A five-factor asset pricing model}.
\newblock \bibinfo{journal}{\emph{Journal of financial economics}}
  \bibinfo{volume}{116}, \bibinfo{number}{1} (\bibinfo{year}{2015}),
  \bibinfo{pages}{1--22}.
\newblock


\bibitem[\protect\citeauthoryear{Feng, Giglio, and Xiu}{Feng
  et~al\mbox{.}}{2020}]%
        {feng2020taming}
\bibfield{author}{\bibinfo{person}{Guanhao Feng}, \bibinfo{person}{Stefano
  Giglio}, {and} \bibinfo{person}{Dacheng Xiu}.}
  \bibinfo{year}{2020}\natexlab{}.
\newblock \showarticletitle{Taming the factor zoo: A test of new factors}.
\newblock \bibinfo{journal}{\emph{The Journal of Finance}}
  \bibinfo{volume}{75}, \bibinfo{number}{3} (\bibinfo{year}{2020}),
  \bibinfo{pages}{1327--1370}.
\newblock


\bibitem[\protect\citeauthoryear{Giglio and Xiu}{Giglio and Xiu}{2017}]%
        {giglio2017inference}
\bibfield{author}{\bibinfo{person}{Stefano Giglio} {and}
  \bibinfo{person}{Dacheng Xiu}.} \bibinfo{year}{2017}\natexlab{}.
\newblock \bibinfo{booktitle}{\emph{Inference on risk premia in the presence of
  omitted factors}}.
\newblock \bibinfo{type}{{T}echnical {R}eport}. \bibinfo{institution}{National
  Bureau of Economic Research}.
\newblock


\bibitem[\protect\citeauthoryear{Harvey and Liu}{Harvey and Liu}{2021}]%
        {HARVEY2021}
\bibfield{author}{\bibinfo{person}{Campbell~R. Harvey} {and}
  \bibinfo{person}{Yan Liu}.} \bibinfo{year}{2021}\natexlab{}.
\newblock \showarticletitle{Lucky factors}.
\newblock \bibinfo{journal}{\emph{Journal of Financial Economics}}
  (\bibinfo{year}{2021}).
\newblock
\showISSN{0304-405X}
\urldef\tempurl%
\url{https://doi.org/10.1016/j.jfineco.2021.04.014}
\showDOI{\tempurl}


\bibitem[\protect\citeauthoryear{Harvey, Liu, and Zhu}{Harvey
  et~al\mbox{.}}{2015}]%
        {harvey2015and}
\bibfield{author}{\bibinfo{person}{Campbell~R Harvey}, \bibinfo{person}{Yan
  Liu}, {and} \bibinfo{person}{Heqing Zhu}.} \bibinfo{year}{2015}\natexlab{}.
\newblock \showarticletitle{… and the cross-section of expected returns}.
\newblock \bibinfo{journal}{\emph{The Review of Financial Studies}}
  \bibinfo{volume}{29}, \bibinfo{number}{1} (\bibinfo{year}{2015}),
  \bibinfo{pages}{5--68}.
\newblock


\bibitem[\protect\citeauthoryear{Heckerman, Geiger, and Chickering}{Heckerman
  et~al\mbox{.}}{1995}]%
        {heckerman1995}
\bibfield{author}{\bibinfo{person}{David Heckerman}, \bibinfo{person}{Dan
  Geiger}, {and} \bibinfo{person}{David~M Chickering}.}
  \bibinfo{year}{1995}\natexlab{}.
\newblock \showarticletitle{Learning Bayesian networks: The combination of
  knowledge and statistical data}.
\newblock \bibinfo{journal}{\emph{Machine learning}} \bibinfo{volume}{20},
  \bibinfo{number}{3} (\bibinfo{year}{1995}), \bibinfo{pages}{197--243}.
\newblock


\bibitem[\protect\citeauthoryear{Himberg, Hyv{\"a}rinen, and Esposito}{Himberg
  et~al\mbox{.}}{2004}]%
        {himberg2004validating}
\bibfield{author}{\bibinfo{person}{Johan Himberg}, \bibinfo{person}{Aapo
  Hyv{\"a}rinen}, {and} \bibinfo{person}{Fabrizio Esposito}.}
  \bibinfo{year}{2004}\natexlab{}.
\newblock \showarticletitle{Validating the independent components of
  neuroimaging time series via clustering and visualization}.
\newblock \bibinfo{journal}{\emph{Neuroimage}} \bibinfo{volume}{22},
  \bibinfo{number}{3} (\bibinfo{year}{2004}), \bibinfo{pages}{1214--1222}.
\newblock


\bibitem[\protect\citeauthoryear{Hou, Xue, and Zhang}{Hou
  et~al\mbox{.}}{2015}]%
        {hou2015digesting}
\bibfield{author}{\bibinfo{person}{Kewei Hou}, \bibinfo{person}{Chen Xue},
  {and} \bibinfo{person}{Lu Zhang}.} \bibinfo{year}{2015}\natexlab{}.
\newblock \showarticletitle{Digesting anomalies: An investment approach}.
\newblock \bibinfo{journal}{\emph{The Review of Financial Studies}}
  \bibinfo{volume}{28}, \bibinfo{number}{3} (\bibinfo{year}{2015}),
  \bibinfo{pages}{650--705}.
\newblock


\bibitem[\protect\citeauthoryear{Hou, Xue, and Zhang}{Hou
  et~al\mbox{.}}{2017}]%
        {hou2017replicating}
\bibfield{author}{\bibinfo{person}{Kewei Hou}, \bibinfo{person}{Chen Xue},
  {and} \bibinfo{person}{Lu Zhang}.} \bibinfo{year}{2017}\natexlab{}.
\newblock \bibinfo{booktitle}{\emph{Replicating Anomalies}}.
\newblock \bibinfo{type}{{T}echnical {R}eport}. \bibinfo{institution}{National
  Bureau of Economic Research}.
\newblock


\bibitem[\protect\citeauthoryear{Hoyer, Janzing, Mooij, Peters, and
  Sch{\"o}lkopf}{Hoyer et~al\mbox{.}}{2008}]%
        {hoyer}
\bibfield{author}{\bibinfo{person}{Patrik Hoyer}, \bibinfo{person}{Dominik
  Janzing}, \bibinfo{person}{Joris~M Mooij}, \bibinfo{person}{Jonas Peters},
  {and} \bibinfo{person}{Bernhard Sch{\"o}lkopf}.}
  \bibinfo{year}{2008}\natexlab{}.
\newblock \showarticletitle{Nonlinear causal discovery with additive noise
  models}.
\newblock \bibinfo{journal}{\emph{Advances in neural information processing
  systems}}  \bibinfo{volume}{21} (\bibinfo{year}{2008}),
  \bibinfo{pages}{689--696}.
\newblock


\bibitem[\protect\citeauthoryear{Huang, Zhang, Lin, Sch{\"o}lkopf, and
  Glymour}{Huang et~al\mbox{.}}{2018}]%
        {huang2018generalized}
\bibfield{author}{\bibinfo{person}{Biwei Huang}, \bibinfo{person}{Kun Zhang},
  \bibinfo{person}{Yizhu Lin}, \bibinfo{person}{Bernhard Sch{\"o}lkopf}, {and}
  \bibinfo{person}{Clark Glymour}.} \bibinfo{year}{2018}\natexlab{}.
\newblock \showarticletitle{Generalized score functions for causal discovery}.
  In \bibinfo{booktitle}{\emph{Proceedings of the 24th ACM SIGKDD International
  Conference on Knowledge Discovery \& Data Mining}}.
  \bibinfo{pages}{1551--1560}.
\newblock


\bibitem[\protect\citeauthoryear{Huang, Zhang, Zhang, Ramsey, Sanchez-Romero,
  Glymour, and Sch{\"o}lkopf}{Huang et~al\mbox{.}}{2020}]%
        {huang2020}
\bibfield{author}{\bibinfo{person}{Biwei Huang}, \bibinfo{person}{Kun Zhang},
  \bibinfo{person}{Jiji Zhang}, \bibinfo{person}{Joseph Ramsey},
  \bibinfo{person}{Ruben Sanchez-Romero}, \bibinfo{person}{Clark Glymour},
  {and} \bibinfo{person}{Bernhard Sch{\"o}lkopf}.}
  \bibinfo{year}{2020}\natexlab{}.
\newblock \showarticletitle{Causal discovery from heterogeneous/nonstationary
  data}.
\newblock \bibinfo{journal}{\emph{Journal of Machine Learning Research}}
  \bibinfo{volume}{21}, \bibinfo{number}{89} (\bibinfo{year}{2020}),
  \bibinfo{pages}{1--53}.
\newblock


\bibitem[\protect\citeauthoryear{Hyvarinen}{Hyvarinen}{1999}]%
        {hyvarinen1999fast}
\bibfield{author}{\bibinfo{person}{Aapo Hyvarinen}.}
  \bibinfo{year}{1999}\natexlab{}.
\newblock \showarticletitle{Fast and robust fixed-point algorithms for
  independent component analysis}.
\newblock \bibinfo{journal}{\emph{IEEE transactions on Neural Networks}}
  \bibinfo{volume}{10}, \bibinfo{number}{3} (\bibinfo{year}{1999}),
  \bibinfo{pages}{626--634}.
\newblock


\bibitem[\protect\citeauthoryear{Hyv{\"a}rinen, Zhang, Shimizu, and
  Hoyer}{Hyv{\"a}rinen et~al\mbox{.}}{2010}]%
        {hyvarinen}
\bibfield{author}{\bibinfo{person}{Aapo Hyv{\"a}rinen}, \bibinfo{person}{Kun
  Zhang}, \bibinfo{person}{Shohei Shimizu}, {and} \bibinfo{person}{Patrik~O
  Hoyer}.} \bibinfo{year}{2010}\natexlab{}.
\newblock \showarticletitle{Estimation of a structural vector autoregression
  model using non-gaussianity.}
\newblock \bibinfo{journal}{\emph{Journal of Machine Learning Research}}
  \bibinfo{volume}{11}, \bibinfo{number}{5} (\bibinfo{year}{2010}).
\newblock


\bibitem[\protect\citeauthoryear{Ilmanen and Kizer}{Ilmanen and Kizer}{2012}]%
        {ilmanen2012death}
\bibfield{author}{\bibinfo{person}{Antti Ilmanen} {and} \bibinfo{person}{Jared
  Kizer}.} \bibinfo{year}{2012}\natexlab{}.
\newblock \showarticletitle{The Death of Diversification Has Been
  GreatlyExaggerated}.
\newblock \bibinfo{journal}{\emph{The Journal of Portfolio Management}}
  \bibinfo{volume}{38}, \bibinfo{number}{3} (\bibinfo{year}{2012}),
  \bibinfo{pages}{15--27}.
\newblock


\bibitem[\protect\citeauthoryear{Jensen, Black, and Scholes}{Jensen
  et~al\mbox{.}}{1972}]%
        {jensen1972capital}
\bibfield{author}{\bibinfo{person}{Michael~C Jensen}, \bibinfo{person}{Fischer
  Black}, {and} \bibinfo{person}{Myron~S Scholes}.}
  \bibinfo{year}{1972}\natexlab{}.
\newblock \showarticletitle{The capital asset pricing model: Some empirical
  tests}.
\newblock In \bibinfo{booktitle}{\emph{Studies in the Theory of Capital
  Markets}}. \bibinfo{publisher}{Praeger Publishers Inc}.
\newblock


\bibitem[\protect\citeauthoryear{Kremer, Talmaciu, and Paterlini}{Kremer
  et~al\mbox{.}}{2018}]%
        {kremer2018risk}
\bibfield{author}{\bibinfo{person}{Philipp~J Kremer}, \bibinfo{person}{Andreea
  Talmaciu}, {and} \bibinfo{person}{Sandra Paterlini}.}
  \bibinfo{year}{2018}\natexlab{}.
\newblock \showarticletitle{Risk minimization in multi-factor portfolios: What
  is the best strategy?}
\newblock \bibinfo{journal}{\emph{Annals of Operations Research}}
  \bibinfo{volume}{266}, \bibinfo{number}{1} (\bibinfo{year}{2018}),
  \bibinfo{pages}{255--291}.
\newblock


\bibitem[\protect\citeauthoryear{McLean and Pontiff}{McLean and
  Pontiff}{2016}]%
        {mclean2016does}
\bibfield{author}{\bibinfo{person}{R~David McLean} {and}
  \bibinfo{person}{Jeffrey Pontiff}.} \bibinfo{year}{2016}\natexlab{}.
\newblock \showarticletitle{Does academic research destroy stock return
  predictability?}
\newblock \bibinfo{journal}{\emph{The Journal of Finance}}
  \bibinfo{volume}{71}, \bibinfo{number}{1} (\bibinfo{year}{2016}),
  \bibinfo{pages}{5--32}.
\newblock


\bibitem[\protect\citeauthoryear{Moneta, Entner, Hoyer, and Coad}{Moneta
  et~al\mbox{.}}{2013}]%
        {moneta2013causal}
\bibfield{author}{\bibinfo{person}{Alessio Moneta}, \bibinfo{person}{Doris
  Entner}, \bibinfo{person}{Patrik~O Hoyer}, {and} \bibinfo{person}{Alex
  Coad}.} \bibinfo{year}{2013}\natexlab{}.
\newblock \showarticletitle{Causal inference by independent component analysis:
  Theory and applications}.
\newblock \bibinfo{journal}{\emph{Oxford Bulletin of Economics and Statistics}}
  \bibinfo{volume}{75}, \bibinfo{number}{5} (\bibinfo{year}{2013}),
  \bibinfo{pages}{705--730}.
\newblock


\bibitem[\protect\citeauthoryear{Pearl}{Pearl}{2009}]%
        {pearl2009causality}
\bibfield{author}{\bibinfo{person}{Judea Pearl}.}
  \bibinfo{year}{2009}\natexlab{}.
\newblock \bibinfo{booktitle}{\emph{Causality}}.
\newblock \bibinfo{publisher}{Cambridge university press}.
\newblock


\bibitem[\protect\citeauthoryear{Peters, Janzing, and Schlkopf}{Peters
  et~al\mbox{.}}{2017a}]%
        {elcainf1}
\bibfield{author}{\bibinfo{person}{Jonas Peters}, \bibinfo{person}{Dominik
  Janzing}, {and} \bibinfo{person}{Bernhard Schlkopf}.}
  \bibinfo{year}{2017}\natexlab{a}.
\newblock \bibinfo{booktitle}{\emph{Elements of Causal Inference: Foundations
  and Learning Algorithms}}.
\newblock \bibinfo{publisher}{The MIT Press}. 33--41 pages.
\newblock
\showISBNx{0262037319}


\bibitem[\protect\citeauthoryear{Peters, Janzing, and Schlkopf}{Peters
  et~al\mbox{.}}{2017b}]%
        {elcainf}
\bibfield{author}{\bibinfo{person}{Jonas Peters}, \bibinfo{person}{Dominik
  Janzing}, {and} \bibinfo{person}{Bernhard Schlkopf}.}
  \bibinfo{year}{2017}\natexlab{b}.
\newblock \bibinfo{booktitle}{\emph{Elements of Causal Inference: Foundations
  and Learning Algorithms}}.
\newblock \bibinfo{publisher}{The MIT Press}. 1--14 pages.
\newblock
\showISBNx{0262037319}


\bibitem[\protect\citeauthoryear{Peters, Mooij, Janzing, and
  Sch{\"o}lkopf}{Peters et~al\mbox{.}}{2014}]%
        {peters2014}
\bibfield{author}{\bibinfo{person}{Jonas Peters}, \bibinfo{person}{Joris~M
  Mooij}, \bibinfo{person}{Dominik Janzing}, {and} \bibinfo{person}{Bernhard
  Sch{\"o}lkopf}.} \bibinfo{year}{2014}\natexlab{}.
\newblock \showarticletitle{Causal discovery with continuous additive noise
  models}.
\newblock \bibinfo{journal}{\emph{Journal of Machine Learning Research}}
  (\bibinfo{year}{2014}).
\newblock


\bibitem[\protect\citeauthoryear{Reichenbach}{Reichenbach}{1956}]%
        {reich}
\bibfield{author}{\bibinfo{person}{Hans Reichenbach}.}
  \bibinfo{year}{1956}\natexlab{}.
\newblock \bibinfo{booktitle}{\emph{The Direction of Time}}.
\newblock \bibinfo{publisher}{University of California Press}.
\newblock


\bibitem[\protect\citeauthoryear{Schwarz et~al\mbox{.}}{Schwarz
  et~al\mbox{.}}{1978}]%
        {schwarz1978estimating}
\bibfield{author}{\bibinfo{person}{Gideon Schwarz} {et~al\mbox{.}}}
  \bibinfo{year}{1978}\natexlab{}.
\newblock \showarticletitle{Estimating the dimension of a model}.
\newblock \bibinfo{journal}{\emph{Annals of statistics}} \bibinfo{volume}{6},
  \bibinfo{number}{2} (\bibinfo{year}{1978}), \bibinfo{pages}{461--464}.
\newblock


\bibitem[\protect\citeauthoryear{Sharpe}{Sharpe}{1964}]%
        {sharpe1964capital}
\bibfield{author}{\bibinfo{person}{William~F Sharpe}.}
  \bibinfo{year}{1964}\natexlab{}.
\newblock \showarticletitle{Capital asset prices: A theory of market
  equilibrium under conditions of risk}.
\newblock \bibinfo{journal}{\emph{The journal of finance}}
  \bibinfo{volume}{19}, \bibinfo{number}{3} (\bibinfo{year}{1964}),
  \bibinfo{pages}{425--442}.
\newblock


\bibitem[\protect\citeauthoryear{Shimizu, Hoyer, Hyv\"{a}rinen, and
  Kerminen}{Shimizu et~al\mbox{.}}{2006}]%
        {lingam}
\bibfield{author}{\bibinfo{person}{Shohei Shimizu}, \bibinfo{person}{Patrik~O.
  Hoyer}, \bibinfo{person}{Aapo Hyv\"{a}rinen}, {and} \bibinfo{person}{Antti
  Kerminen}.} \bibinfo{year}{2006}\natexlab{}.
\newblock \showarticletitle{A Linear Non-Gaussian Acyclic Model for Causal
  Discovery}.
\newblock \bibinfo{journal}{\emph{J. Mach. Learn. Res.}}  \bibinfo{volume}{7}
  (\bibinfo{date}{Dec.} \bibinfo{year}{2006}), \bibinfo{pages}{2003–2030}.
\newblock
\showISSN{1532-4435}


\bibitem[\protect\citeauthoryear{Shimizu, Inazumi, Sogawa, Hyv{\"a}rinen,
  Kawahara, Washio, Hoyer, and Bollen}{Shimizu et~al\mbox{.}}{2011}]%
        {shimizu2011directlingam}
\bibfield{author}{\bibinfo{person}{Shohei Shimizu}, \bibinfo{person}{Takanori
  Inazumi}, \bibinfo{person}{Yasuhiro Sogawa}, \bibinfo{person}{Aapo
  Hyv{\"a}rinen}, \bibinfo{person}{Yoshinobu Kawahara},
  \bibinfo{person}{Takashi Washio}, \bibinfo{person}{Patrik~O Hoyer}, {and}
  \bibinfo{person}{Kenneth Bollen}.} \bibinfo{year}{2011}\natexlab{}.
\newblock \showarticletitle{DirectLiNGAM: A direct method for learning a linear
  non-Gaussian structural equation model}.
\newblock \bibinfo{journal}{\emph{The Journal of Machine Learning Research}}
  \bibinfo{volume}{12} (\bibinfo{year}{2011}), \bibinfo{pages}{1225--1248}.
\newblock


\bibitem[\protect\citeauthoryear{Sims}{Sims}{1980}]%
        {sims1980macroeconomics}
\bibfield{author}{\bibinfo{person}{Christopher~A Sims}.}
  \bibinfo{year}{1980}\natexlab{}.
\newblock \showarticletitle{Macroeconomics and reality}.
\newblock \bibinfo{journal}{\emph{Econometrica: journal of the Econometric
  Society}} (\bibinfo{year}{1980}), \bibinfo{pages}{1--48}.
\newblock


\bibitem[\protect\citeauthoryear{Spirtes, Glymour, Scheines, and
  Heckerman}{Spirtes et~al\mbox{.}}{2000}]%
        {spirtes2000}
\bibfield{author}{\bibinfo{person}{Peter Spirtes}, \bibinfo{person}{Clark~N
  Glymour}, \bibinfo{person}{Richard Scheines}, {and} \bibinfo{person}{David
  Heckerman}.} \bibinfo{year}{2000}\natexlab{}.
\newblock \bibinfo{booktitle}{\emph{Causation, prediction, and search}}.
\newblock \bibinfo{publisher}{MIT press}.
\newblock


\bibitem[\protect\citeauthoryear{Vowels, Camgoz, and Bowden}{Vowels
  et~al\mbox{.}}{2021}]%
        {vowels2021d}
\bibfield{author}{\bibinfo{person}{Matthew~J Vowels},
  \bibinfo{person}{Necati~Cihan Camgoz}, {and} \bibinfo{person}{Richard
  Bowden}.} \bibinfo{year}{2021}\natexlab{}.
\newblock \showarticletitle{D'ya like DAGs? A Survey on Structure Learning and
  Causal Discovery}.
\newblock \bibinfo{journal}{\emph{arXiv preprint arXiv:2103.02582}}
  (\bibinfo{year}{2021}).
\newblock


\end{thebibliography}

\end{document}